\documentclass[a4paper,11pt]{article}
\usepackage{pos}
\usepackage{booktabs}

\newcommand{\ssteff}{\ensuremath{\sin^2\theta_{W,\ell}^\text{eff.}}}

\newcommand{\diff}{\ensuremath{\text{d}}}

\newcommand{\tcs}{\ensuremath{\theta_{CS}}}

\title{ Status of the W-boson mass averaging project }
\ShortTitle{ Status of the W-boson mass averaging project }
\author*{Simone Amoroso, on behalf of the Tevatron/LHC W-mass combination Working Group}
\renewcommand{\printHeadAuthors}{Simone Amoroso}
\affiliation{Deutsches Elektronen-Synchrotron (DESY),\\
  Notkestrasse 85, Hamburg, Germany}
\emailAdd{simone.amoroso@desy.de}

\abstract{
We present the current status of the  W-boson mass averaging project, an ongoing effort aimed at combining Tevatron and LHC measurements.
Methods are presented to accurately evaluate the effect of PDFs and other modelling variations on existing measurements.
Based on this approach, the measurements can be corrected to a common modelling reference and to the same PDFs, 
and subsequently combined accounting for PDF correlations in a quantitative way.
We  discuss the combination procedure, and the impact of improvements in the theoretical description of W-boson production and decay.
}

\FullConference{%
  41st International Conference on High Energy physics - ICHEP2022\\
  6-13 July, 2022\\
  Bologna, Italy
}

\begin{document}


\maketitle

\section{Introduction}
The W-boson mass, $m_W$, is a fundamental parameter of the Standard Model,
and its value one of the most important inputs to global electroweak fits~\cite{Erler:2019hds}.
The W-boson mass has been precisely measured at hadron colliders  by the 
CDF~\cite{CDF:2022hxs} and D0~\cite{D0:2013jba} Collaborations at the Tevatron
and by the ATLAS~\cite{ATLAS:2017rzl} and LHCb~\cite{LHCb:2021bjt} Collaborations at the LHC.
In this contribution, we report on the current status of the Tevatron/LHC W-boson mass combination effort,
aiming to provide a world average of $m_W$ determinations endorsed by the experimental collaborations.
Such a combination is particularly motivated in light of the discrepancy between the most recent CDF measurement 
and the   results from the other experiments.

At hadron colliders, measurements of $m_W$  rely on template fits to  kinematic peaks of distributions determined from the leptonic decays of the W boson,
such as the lepton transverse momentum, $p_T^{l}$, or the W-boson transverse mass  $m_T^W$ . 
These final-state distributions carry information about the decaying particle mass, but are also affected
by the description of  W-boson production and decay, such as the rapidity and transverse momentum distributions and its polarization. 
Predictions of these observables are obtained using Monte Carlo (MC) event generators with input parton distribution functions (PDFs).
With each measurement relying on a different generator and PDF choice, 
prior to combining the measurements a coherent theoretical treatment is required to estimate uncertainty correlations.
These modelling differences further motivate small adjustments  to the measured values or uncertainties,
an effect which has been neglected in past combinations~\cite{2134139}.

Beyond the interest of improving the overall measurement precision on $m_W$, through this combination effort 
we establish a methodology to combine  present and upcoming measurements of $m_W$
and enable possible updates of the physics  modelling (e.g. PDFs, $p_T^W$) as our theoretical knowledge improves.
In addition, as more electroweak measurements become dominated by PDF uncertainties (i.e. $m_W$ and  $\ssteff$), 
this methodology will permit to consistently correlate them in global electroweak fits.

\section{Combination strategy}

For a consistent combination of different measurements, their correlations need to be determined.
The main source of correlations among the $m_W$ measurements comes from the knowledge of  the proton structure, embedded in the PDFs.
Due to the different center-of-mass energies  and initial states ($pp$ vs $p\bar{p}$) at which the measurements are performed this correlation is however non-trivial.
Uncertainties related to the $p_T^W$ distribution, while significant in size, are evaluated
independently in each experiment through a detailed analysis of Z-boson production, 
and are hence assumed to be uncorrelated.
Other sources of theoretical uncertainties, such  as electroweak corrections, 
are typically very small, and their detailed correlation knowledge do not affect the combination.
Experimental uncertainties are expected to be uncorrelated between experiments.

The combination of $m_W$ results is performed in a  two-step procedure.
At first, the results have to be translated to a common reference model.
The full procedure to correct the published $m_W$ values is decomposed into QCD and PDF effects, 
and allows to improve existing experimental results to include progress in theoretical prediction and PDF determinations.
The published $m_W$ values $m_W^{\rm{published}}$ of each experiment is subject to the correction: 
\begin{equation}
m_W^{\rm{updated}} = m_W^{\rm{published}}-\delta m_W^{\rm{QCD}}-\delta m_W^{\rm{PDF}},
\end{equation}
where $\delta m_W^{\rm{QCD}}$ incorporates any eventual corrections to the QCD modelling 
beyond the originally quoted uncertainties, 
and $\delta m_W^{\rm{PDF}}$  brings all measurements to a common PDF set.
The results such obtained are then combined properly across experiments including correlations.
The model dependence of the result is evaluated by repeating 
this procedure for a relevant set of PDF determinations.

\section{Detector Emulation}
The original $m_W$ measurements have been performed at detector level, 
and reproducing the analyses and detector simulations used by the experiments is a challenging task.
Instead, we make use of a parametrized detector response, tuned on publicly available information (lepton energy/momentum scale and resolution, efficiencies; recoil response) 
and applied on large Monte Carlo event samples at particle-level, generated with varying QCD modelling assumptions and PDF sets.  
This approach has been found sufficient for a reliable evaluation of variations in the underlying generator-level distributions, 
such as PDF uncertainties and extrapolations, and effects in the lepton angular distributions.
After event selection it can reproduce published distributions at the \% level, 
with systematic uncertainty on $\delta m_W$  estimated to be at the level of 1-2 MeV.

\section{QCD modelling aspects}

We concentrate here on the description of the angular distribution of the decay leptons.
After integrating over azimuthal angle, we can express the W-boson cross-section to all orders in the strong coupling as:
\begin{equation}
    \label{eq:differential angular Drell-Yan cross section at NLOQCD}
    \diff\sigma=\sigma_\text{unpol}\biggr[(1+\cos^2\tcs)+\frac{A_0}{2}(1-3\cos^2\tcs)+A_4\cos\tcs\biggr],
\end{equation}
with $\sigma_\text{unpol}$ the unpolarised cross-section, $\cos\tcs$ the polar angle in the Collins-Soper frame~\cite{Collins:1977iv},
and $A_0$ and $A_4$ angular coefficients which will in general depend on the boson kinematics. 
Theoretical predictions at fixed-order for the angular coefficients are currently known  up to $O(\alpha_S^3)$~\cite{Gauld:2017tww,Pellen:2022fom},
and describe to a very good degree of accuracy the measured coefficients at the LHC~\cite{ATLAS:2016rnf}.

Different theoretical predictions have been used for the various $m_W$ measurements,
reflecting the theoretical state of the art at the time of their preparation.
For the descriptions of the W-boson production and decay, CDF used events generated with the Resbos-C~\cite{Ladinsky:1993zn} code at NLO+NLL  
while the Resbos-CP~\cite{Balazs:1997xd} code  at NNLO+NNLL  has been used by  D0.
ATLAS and LHCb relied on  parton shower generated events from Pythia8~\cite{Bierlich:2022pfr},
but  reweighted  the angular distributions to $O(\alpha_S^2)$ fixed-order predictions from DYNNLO~\cite{Catani:2009sm}.

We generated event samples using the same  MC generation chain as the original measurements.
In addition,  MC samples have been generated using Powheg Z\_{EW}~\cite{Barze:2013fru}, 
MiNNLOPS~\cite{Monni:2019whf,Monni:2020nks} and  an updated version of Resbos (dubbed here Resbos2)
with an improved treatment of spin correlations~\cite{Isaacson:2022rts}.
This allows to study the impact of computing $\delta m_W^{\rm{QCD}}$ and $\delta m_W^{\rm{PDF}}$ with predictions at different theoretical accuracies.

\begin{figure}[htbp]
\centering
\includegraphics[width=0.48\textwidth]{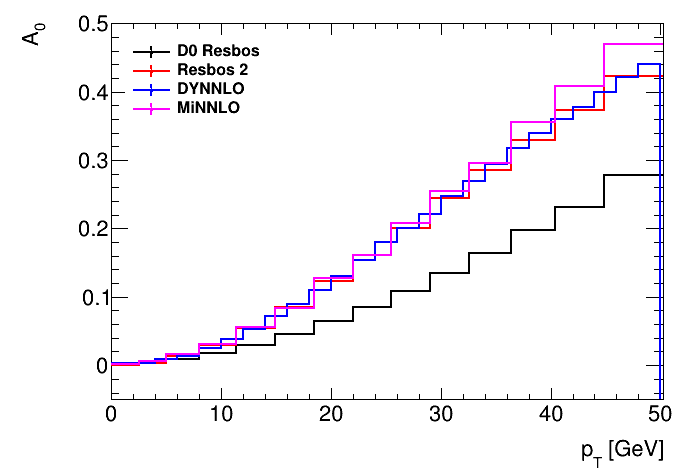}
\includegraphics[width=0.48\textwidth]{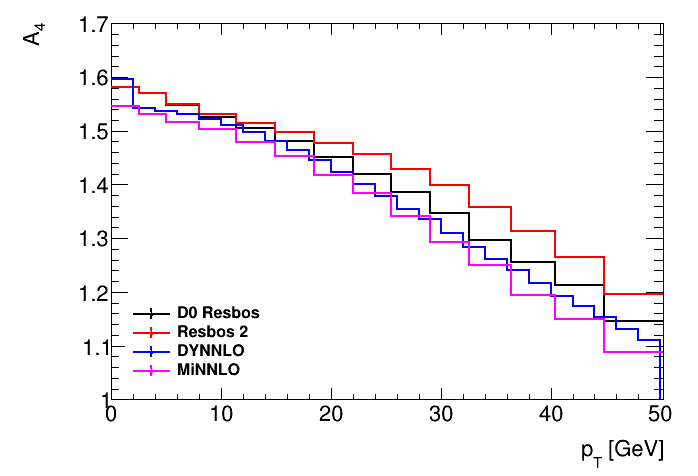}
\caption{The angular coefficients $A_0$ (left plot) and $A_4$ (right plot) in W-boson production shown as a function of $p_T^W$ in $p\bar{p}$ collisions at $\sqrt{s}=1.96$~TeV.
The predictions from the legacy D0 version of Resbos (black) are compared to predictions from an updated version of Resbos (red), fixed-order predictions at $O(\alpha_S)$ from DYNNLO (blue)
and MC predictions at $O(\alpha_S^2)$ from MiNNLOPS (magenta).}
\label{fig:ai}
\end{figure}

We compare in Fig.~\ref{fig:ai} different predictions for $A_0$, $A_4$ a function of $p_T^W$.
Large differences can be seen between the Resbos-CP version employed by D0 and the modern event generator or fixed-order codes. 
Similar differences are also observed with respect to the Resbos-C version used by CDF.
This is explained as  the legacy Resbos codes do not perform a consistent resummation of all the angular coefficients,
but only  $A_4$ and $\sigma_\text{unpol}$ receive corrections from the resummation.
This behavior has been corrected in ResBos2, which largely reproduces the fixed-order behavior.

\begin{figure}[htbp]
\centering
\includegraphics[width=0.48\textwidth]{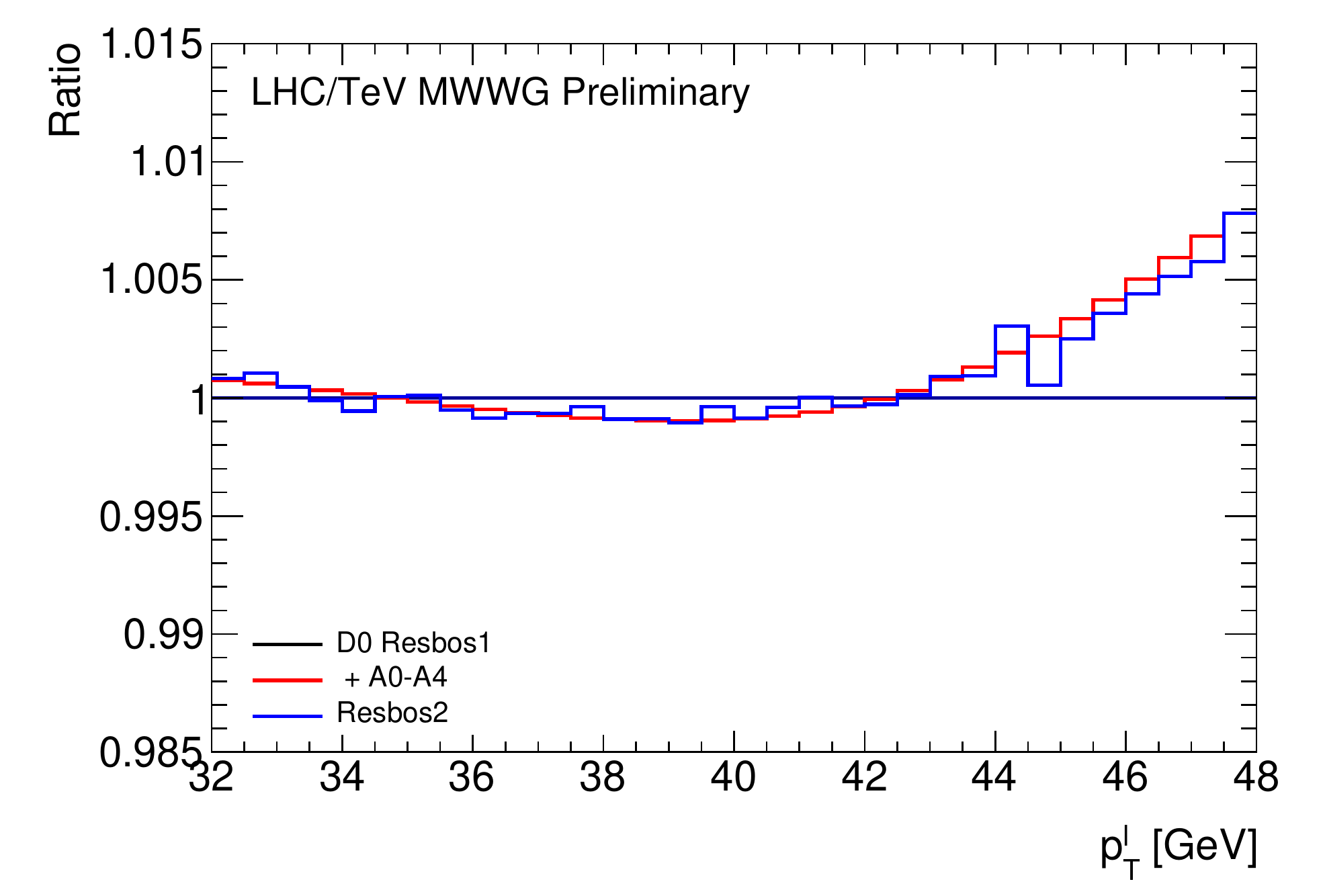}
\includegraphics[width=0.48\textwidth]{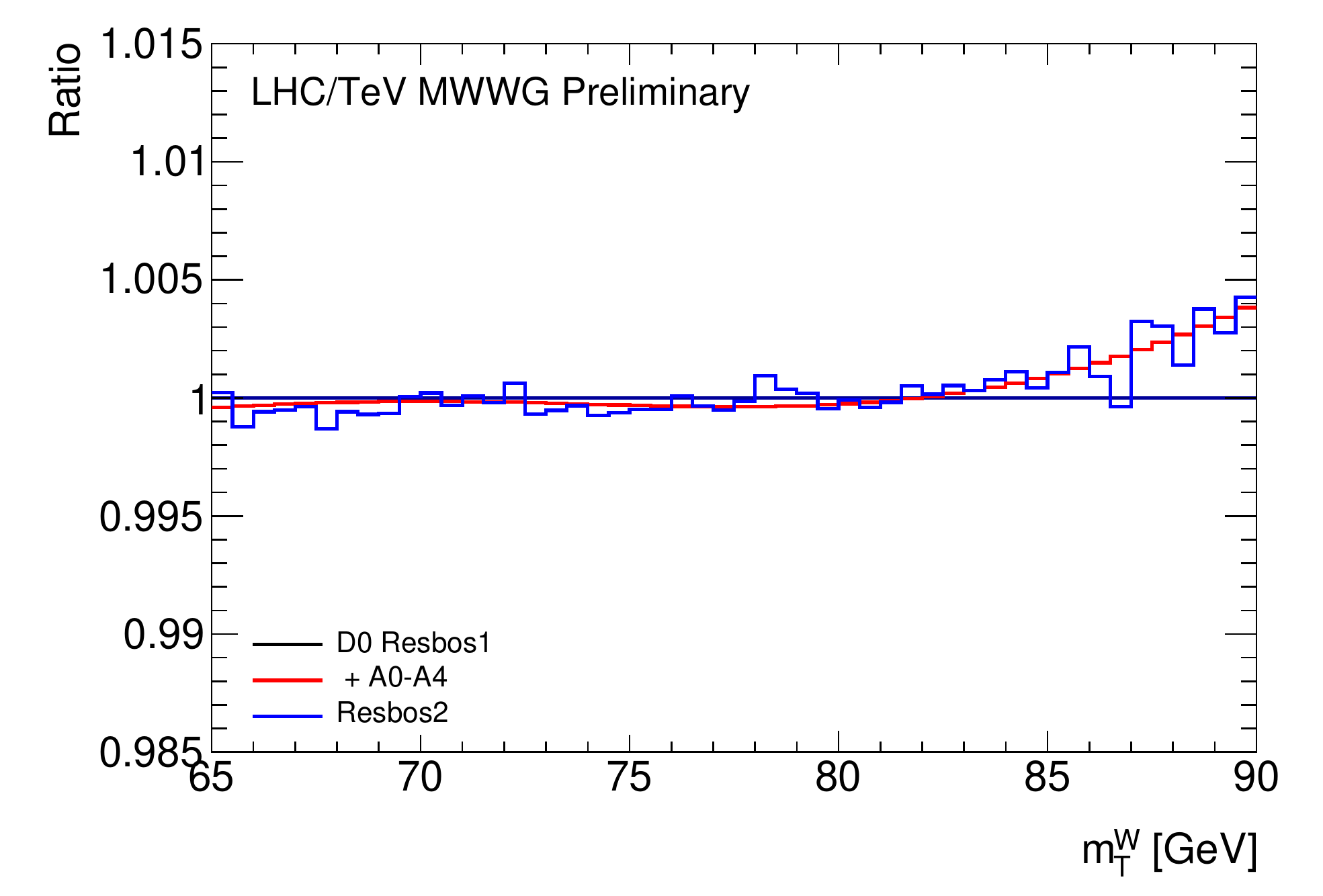}
\caption{Relative effect of the generator corrections $\delta m_W^{\rm{QCD}}$ 
on the lepton transverse momentum (left plot) and W-boson transverse mass distributions (right plot),
after including  detector effect and  applying the $p_T^W$ constraint.
The correction is obtained by reweighting the angular coefficients to the Resbos 2 predictions (red) and shown with respect to the D0 legacy Resbos-CP code (black) 
and compared to the directly generated Resbos2 predictions (blue).
}
\label{fig:pT}

\end{figure}

\begin{table}[htbp]
   \centering
   \begin{tabular}{lcccccccc}
     \toprule
     Correction     & & \multicolumn{7}{c}{$\delta m_W^{\mathrm{QCD}}$ [MeV]}\\
     & & \multicolumn{3}{c}{$p_T^W$-constrained} & & \multicolumn{3}{c}{No constraint}\\
     & & $p_T^{l}$ & $m_T$ & $p_T^{\nu}$ & & $p_T^{l}$  & $m_T$ & $p_T^{\nu}$  \\
     \midrule
     Invariant mass & & $<0.1$ & $<0.1$ & $<0.1$ & & $<0.1$ & $<0.1$ & $<0.1$ \\
     Rapidity       & & $<0.1$ & $<0.1$ & $<0.1$ & & $<0.1$ & $<0.1$ & $<0.1$ \\
     \midrule
     $A_0$          & &     7.6 &   10.0 &   15.8  & &   16.0 &   12.6 &   19.5 \\
     $A_1$          & &    -2.4 &   -1.9 &   -1.8  & &   -1.2 &   -1.6 &   -1.4 \\
     $A_2$          & &    -3.0 &   -2.6 &    2.9  & &   -4.2 &   -3.0 &    2.3 \\
     $A_3$          & &     2.9 &    1.6 &   -0.5  & &    3.5 &    1.8 &   -0.2 \\
     $A_4$          & &     2.4 &   -0.1 &   -0.5  & &    0.1 &   -0.7 &   -1.0 \\
     $A_0-A_4$      & &     7.6 &    7.0 &   16.0  & &   14.1 &    9.1 &   18.9 \\
     \midrule
     Total         & &      7.6 &    7.0 &   16.0 & &   14.1 &    9.1 &   18.9 \\
     \midrule
     \textsc{ResBos2}      & & 7.3$\pm$1.1 & 8.4$\pm$1.0 & 16.6$\pm$1.2 & & 13.9$\pm$1.1 & 10.3$\pm$1.0 & 19.8$\pm$1.2 \\
     \midrule
     Non-closure   & & -0.3$\pm$1.1 & 1.4$\pm$1.0 & 0.6$\pm$1.2 & & -0.2$\pm$1.1 & 1.2$\pm$1.0 & 0.9$\pm$1.2 \\
     \bottomrule
   \end{tabular}
   \caption{Impact of  the angular coefficients reweighting in the D0 \textsc{ResBos-CP} events to those of \textsc{ResBos2}, 
   compared to  a direct fit of \textsc{ResBos1} to \textsc{ResBos2}. Results are shown for the $p_T^{l}$,  $p_T^{\nu}$ and $m_T^{W}$ distributions with and without a constraint to leave the $p_T^{W}$ unchanged. \label{tab:updatesD0}}
\end{table}

We evaluated a correction by reweighting the angular coefficients in the original D0 Resbos-CP sample to the $O(\alpha_S)$ prediction of the Resbos2 sample, 
and compare it with the shift obtained by directly fitting the Resbos2 distributions.
The impact of this correction for the D0 measurement is shown in Tab.~\ref{tab:updatesD0}. 
When constraining the $p_T^W$ distribution to stay unchanged under the  corrections
yields shifts of $7.3\pm1.1$~MeV and $8.4\pm1.0$~MeV  for the $p_T^l$ and $m_T$ fits. 
If the $p_T^W$ distribution is instead left unconstrained, shifts of $13.9\pm1.1$~MeV  and $10.3\pm1.0$~MeV  
for the  $p_T^l$ and $m_T$ fits are obtained. 
As the measurements constrain $p_T^W$ only through the measured $p_T^Z$ distribution,
these numbers are to be considered as an upper bound of the effect.

\section{Summary}
We presented studies towards a first combination of Tevatron and LHC measurements of $m_W$. 
Measurement correlations are dominated by PDF uncertainties and are evaluated through a simpliﬁed emulation of the detector response.
Extensive comparisons against state-of-the-art QCD predictions highlighted the need to correct the Tevatron
$m_W$ determinations for the treatment of lepton angular distributions in W-boson decays in the
legacy Resbos codes. In the context of the D0 measurement, correcting the spin correlations affects the
result at the level of -10~MeV, depending on the distribution used in the fit.

\bibliographystyle{JHEP}
\bibliography{wcombi}
\end{document}